\documentclass[preprint,showpacs,preprintnumbers,amsmath,amssymb]{revtex4}
\usepackage{graphicx}
\usepackage{epsf}
\usepackage{dcolumn}
\usepackage{bm}

\begin{document}

\title{~~~~\\ Consistency in NLO analyses of inclusive  and \\  semi-inclusive polarized DIS data.\\ ~~~~~~~ }
\author{\large  G. A. Navarro, R. Sassot\footnote{Partially supported by Conicet and Fundaci\'on Antorchas} }  
\email{sassot@df.uba.ar}
\affiliation{
 Departamento de F\'\i sica, Facultad de Ciencias Exactas y Naturales \\
 Universidad de Buenos Aires, Pabell\'on I, Ciudad Universitaria \\ 
                       (1428) Buenos Aires, Argentina }
\date{\today }

\begin{abstract}
We perform a detailed study of the consistency between different sets of polarized 
deep inelastic scattering data and theory, from the standpoint of a next to leading order QCD 
global analysis, and following the criteria proposed by Collins and Pumplin. In face 
of recent suggestions that challenge the usual assumption about parent parton spin 
independence of unpolarized fragmentation functions, we specially focus on polarized 
semi-inclusive data.
\end{abstract}

\maketitle

\section{Introduction}

Ever since the renewed burst of interest in the spin structure of the nucleon, triggered
fifteen years ago by the measurement of the proton spin dependent deep inelastic 
scattering structure function $g_1^p$ by the EMC experiment \cite{EMC}, polarized 
deep inelastic scattering has evolved into a very prolific field for combined theoretical
and experimental efforts \cite{review}. As result of this activity, an increasingly 
precise QCD improved partonic description of polarized nucleons has emerged, phrased 
in terms of more refined and exhaustive extractions of polarized parton distributions.
       
Indeed, the number of QCD analyses of polarized inclusive deep inelastic scattering (DIS)
data at next to leading order (NLO) accuracy has rapidly grown in the few last years 
\cite{fits}. In some cases these studies also take into account polarized semi-inclusive deep 
inelastic scattering (SIDIS) data \cite{DSS,DS,SW}. The usual outcome from these analyses
are different sets of parton distributions which reproduce fairly well most data sets 
within the quoted errors, with overall $\chi^2$ values pretty close to the number $N$ of 
degrees of freedom (d.o.f) as required by the familiar `hypothesis-testing criterion'. 

Although the apparent fulfillment of this requirement, the resulting parton distributions not only 
suffer from large uncertainties, but in some cases small subsets of inclusive data are 
badly fitted. This facts point to the need of a more stringent criterion for assessing the goodness 
of a particular fit, and also the compatibility between data sets.

In the case of polarized semi-inclusive data, the motivation for a more careful
analysis is twofold. For one side, inclusive data do not provide enough information 
for a complete separation of the quark and antiquark distributions of  different 
flavors, so the analyses must rely either on SIDIS data or on some external assumption.

From another side, the standard
procedure used to analyze semi-inclusive data, in terms of polarized parton 
densities and unpolarized fragmentation functions, has been criticized 
questioning, in first place, the accuracy of the present generation of 
fragmentation functions \cite{ELLIOT}, suggesting that the 
usual assumptions about spin independence of fragmentation process  may not 
hold \cite{GR}, and also addressing the issue of target fragmentation contributions \cite{Kos}.  

Regarding the first issue, recently there has been an increasing amount activity around the 
issue of the polarization of sea quarks in the proton \cite{mod}. This interest
has been driven in part by the confirmation of the isospin symmetry breaking
at sea quark level in unpolarized DIS \cite{unpol}, by the level of refinement 
attained in QCD global analysis of spin dependent data \cite{fits}, and by the 
expected precision of forthcoming polarized semi-inclusive measurements, 
polarized Drell Yan dilepton production, and prompt photon production 
\cite{future}.

Moreover, in spite of the encouraging achievements in polarized semi-inclusive 
measurements in the few last years \cite{SMC,HERMES}, which have shown 
to be in good agreement with inclusive data \cite{DSS}, the available 
data fail to constrain unambiguously the polarization of sea quarks in the 
proton when included in global QCD fits performed following just the
hypothesis testing criterion \cite{DS}. 

In face of this situation, it is worthwhile assessing the goodness of global 
fits to polarized data, involving both inclusive semi-inclusive measurements, with more
stringent criteria, in the line of what has been proposed in reference \cite{COLLINS},
and applied to LO unpolarized sets of data in \cite{IJ}. The main point of 
this approach is to apply the parameter fitting criterion to subsets of data
in the global fit, assessing in this way features as the degree of 
compatibility between subsets of data, the impact of specific data in certain
parameters of the fit, and the overall consistency of the global fit. 
The occurrence of such an inconsistency may hint, besides an unexpected error source in 
the experiment, the non validity of assumptions or other inputs of the theoretical
calculation.

In the following, after defining our conventions and sketching the computation of the relevant
observables, we present a global fit of polarized inclusive and 
semi-inclusive data exploring the dependence of the overall $\chi^2_{tot}$ 
of the fit on the respective $\chi^2_i$ for each particular experiment by means of the 
Lagrange multiplier method \cite{LM}. 
The analysis is performed in two stages. In the first one we restrict the analysis 
to inclusive data, and to those parton densities which can be extracted there.
In the second stage we add semi-inclusive observables and discriminate between 
valence and sea quark densities of different flavors, but at variance with
reference \cite{DS}, where semi-inclusive data was only allowed to fix the sea quark
polarization, here we leave all the distributions free.

In order to circumvent the extremely time-consuming convolutions integrals 
characteristic of the semi-inclusive observables at NLO, we apply the double 
Mellin transform approach, developed recently in \cite{SW}. This new approach has produced 
results in complete agreement with those obtained with the convolution method, 
but obtained considerably faster.

As result of our analysis, we find that the parameter fitting criterion shows a reasonably good level of internal consistency between the inclusive measurements. 
The addition of SIDIS data sets leads to global fits with $\chi^2/d.o.f.$ values
fairly close to unity, however the increase of the statistical weight of this data
in the fits shows rather different consequences when either positively ($\sigma^{h+}$) or 
negatively ($\sigma^{h-}$) charged final state hadron data is considered. For $\sigma^{h-}$ 
data, the 
increase in weight does not modify significantly those parton distributions coming from 
inclusive sets, suggesting consistency between the different data sets and also between 
data and the theoretical framework. 
However, the analysis of $\sigma^{h+}$ data shows a sizable degree of conflict with the 
rest of the fit.

\section{ Framework.}

Throughout the present analysis, we follow the same conventions and definitions
for the polarized inclusive asymmetries and parton densities as in reference 
\cite{DS}. 
In the totally inclusive case, the spin dependent asymmetries are given by \cite{review}
\begin{widetext}
\begin{eqnarray}
A_1^N(x,Q^2)=
\frac{g_1^N(x,Q^2)}{F_1^N(x,Q^2)}\,=\,\frac{g_1^N(x,Q^2)}{F_2^N(x,Q^2)/{ 2x[ 1+R^N(x,Q^2)] } },
\end{eqnarray}
where the inclusive spin-dependent nucleon structure function $g_1^N(x,Q^2)$ can be written at NLO as a convolution between polarized parton densities for quarks and gluons, $\Delta q_i(x,Q^2)$ and $\Delta g(x,Q^2)$, respectively, and coefficient functions $\Delta C_i(x)$\cite{inclusiva}
\begin{eqnarray}
g_1^N(x,Q^2)=\frac{1}{2}\sum_{q,\bar{q}}e_q^2 \, 
\Bigg[ \Delta q(x,Q^2)\, +\,\frac{\alpha_s(Q^2)}{2\pi} \int_x^1 \frac{dz}{z} \Bigg\{\Delta C_q(z) \Delta q(\frac{x}{z},Q^2)\nonumber \\
+\, \Delta C_g(z) \Delta g(\frac{x}{z},Q^2)\Bigg\}\Bigg].
\end{eqnarray}

A more detailed discussion about these coefficient functions and their factorization scheme dependence can be found in Ref.\cite{newgr}. $F_1^N(x,Q^2)$ is the unpolarized nucleon structure function that can be written in terms of $F_2^N(x,Q^2)$ and R, the ratio of the longitudinal to transverse cross section \cite{review}.

Analogously, for the semi-inclusive asymmetries we have:
\begin{eqnarray}
A_1^{Nh}(x,Q^2)\mid_Z\, \simeq \,\frac{\int_Z \,dz\, g_1^{Nh}(x,z,Q^2)}{\int_Z \,dz\, F_1^{Nh}(x,z,Q^2)},
\end{eqnarray}
bf where the superscript $h$ denotes the hadron detected in the final state, and the variable $z$ is given by the ratio between the hadron energy and that of the spectators in the target. The region $Z$, over which $z$ is integrated, is determined by kinematical cuts applied when measuring the asymmetries.

For the spin dependent structure 
function $g^N_1(x,Q^2)$, we use the NLO expression \cite{NPB}
\begin{eqnarray}
 g_{1}^{N\,h}(x,z,Q) =
%
%
\frac{1}{2}\sum_{q,\bar{q}} \!\!\!\!\!\ e_q^2 \;
\Bigg[ \Delta q \left(x,Q^2\right) D_q^H\left(z,Q^2\right) +
\frac{\alpha_s(Q^2)}{2\pi}
\int_x^1 \frac{d\hat{x}}{\hat{x}} \int_z^1 \frac{d\hat{z}}{\hat{z}}
\Bigg\{
\Delta q \left(\frac{x}{\hat{x}},Q^2\right)
\nonumber \\
\Delta C_{qq}^{(1)}(\hat{x},\hat{z},Q^2)
D_{q}^H\left(\frac{z}{\hat{z}},Q^2\right)+ 
\Delta q \left(\frac{x}{\hat{x}},Q^2\right)
\Delta C_{gq}^{(1)} (\hat{x},\hat{z},Q^2)
\nonumber \\
D_{g}^H\left(\frac{z}{\hat{z}},Q^2\right) +
\Delta {g} \left(\frac{x}{\hat{x}},Q^2\right)
\Delta C_{qg}^{(1)} (\hat{x},\hat{z},Q^2)
D_{q}^H\left(\frac{z}{\hat{z}},Q^2\right)\Bigg\}\Bigg],
\end{eqnarray}
and in order to avoid the convolution integrals in $\hat{x}$ and $\hat{z}$
we switch to moment space in both variables as suggested in \cite{SW}. 
In moment space, the convolution integrals reduce products of the Mellin
moments of the parton densities
\begin{equation}
\Delta f_i^n(Q^2) \equiv \int_0^1 dx \;x^{n-1}  \Delta f_i(x,Q^2) \; ,
\end{equation}
fragmentation functions
\begin{equation}
\Delta D_i^{h\,m}(Q^2) \equiv \int_0^1 dx \;x^{m-1}  \Delta D_i^h(z,Q^2) \; , 
\end{equation}
and the double Mellin transform of the coefficient functions 
$\Delta C_{ij}^{(1)}(x,z,Q^2)$, defined by
\begin{equation}
\Delta C_{ij}^{(1),nm}(Q^2) \equiv
\int_0^1 dx\, x^{n-1} \int_0^1 dz\, z^{m-1} 
\Delta C_{ij}^{(1)}(x,z,Q^2).
\end{equation}
These coefficients can be written as \cite{DSV}

\begin{eqnarray} 
\Delta C_{qq}^{(1),nm}(Q^2)&=& 
C_F \Bigg[
-8-\frac{1}{m^2} +\frac{2}{(m+1)^2}+\frac{1}{n^2}+
\frac{(1+m+n)^2-1}{m (m+1)n(n+1)} \nonumber \\   
&& 
+ \left[ S_1(m) + S_1(n) \right] 
\left\{ 
S_1(m) + S_1(n) -\frac{1}{m(m+1)}-\frac{1}{n(n+1)}\right\} 
\nonumber \\  
&& 
+3 S_2(m) - S_2(n) \Bigg] 
\end{eqnarray}

\begin{eqnarray}
\Delta C_{gq}^{(1),nm} (Q^2)&=& 
C_F \Bigg[
\frac{2-2 m-9 m^2+m^3-m^4+m^5}{m^2 (m-1)^2(m+1)^2}+
\frac{2m}{n (m+1)(m-1)} \nonumber \\  
&& 
-\frac{2-m+m^2}{m (m+1)(m-1)(n+1)}-
\frac{2+m+m^2}{m (m+1)(m-1)} 
\left[ S_1(m) + S_1(n) \right]\nonumber \\ 
&&- \frac{2}{(m+1)n (n+1)} \Bigg]  \\
\Delta C_{qg}^{(1),nm} (Q^2)&=&
%
T_R \frac{n-1}{n(n+1)}
\Bigg[\frac{1}{m-1}-\frac{1}{m}+\frac{1}{n}-S_1(m) - S_1(n)
 \Bigg] \; ,
\end{eqnarray}
\end{widetext}
where we have set the factorization and renormalization scales to $Q^2$. As 
usual, $C_F=4/3$, $T_R=1/2$, and 
\begin{equation}
S_i(n) \equiv \sum_{j=1}^n \frac{1}{j^i}.
\end{equation}

For $u$ and $d$ quarks {\em plus} antiquarks densities at the 
initial scale $Q^2_0 =0.5$ GeV$^2$ we propose
\begin{eqnarray}
 x (\Delta q+\Delta \overline{q})  =  N_{q}
\frac{x^{\alpha_q}(1-x)^{\beta_q}(1+\gamma_q\,
x^{\delta_q})}{B(\alpha_q+1,\beta_q+1)+\gamma_q \, B(\alpha_q+\delta_q+1,\beta_q+1) },\nonumber
\end{eqnarray}
where $B(\alpha,\beta)$ is the standard beta function,
while for strange quarks {\em plus} antiquarks we use
\begin{equation}
 x(\Delta s + \Delta \overline{s}) = 2 N_{s}
\frac{x^{\alpha_{{s}}}(1-x)^{\beta_{{s}}}}
{B(\alpha_{{s}}+1,\beta_{{s}}+1)},
\end{equation}
with a similar parametric form for gluons
\begin{equation}
 x\Delta g =N_{g}
\frac{x^{\alpha_{{g}}}(1-x)^{\beta_{{g}}}}
{B(\alpha_{{g}}+1,\beta_{{g}}+1)}.
\end{equation}

The first moments of the quark densities $\delta q$ ($N_q$) are often related
to the hyperon beta decay constants $F$ and $D$ through the SU(3) symmetry 
relations
\begin{eqnarray} 
\delta u +\delta \overline{u}- \delta d -\delta
\overline{d} & \equiv & N_{u} - N_{d} \nonumber \\
& = & F+D = 1.2573
\end{eqnarray} 
\begin{eqnarray} 
\delta u +\delta
\overline{u}+ \delta d +\delta \overline{d}- 2(\delta s +\delta
\overline{s}) & \equiv & N_{u} + N_{d}  - 4 N_{s}\nonumber \\
 = 3F-D & = & 0.579.
\end{eqnarray}
Under such an assumption, the previous equations would strongly constrain the normalization of the
 quark densities. However, as we are not interested in forcing flavor symmetry, we leave aside 
that strong assumption and  relax the symmetry relations introducing two parameters, $\epsilon_{Bj}$ 
and $\epsilon_{SU(3)}$ respectively. These parameters account quantitatively for eventual departures 
from  flavor symmetry considerations, including also some uncertainties on the low-$x$ behavior, and 
 higher order corrections,
\begin{equation} 
\label{Bj}
N_{u} - N_{d} = (F+D)(1+\epsilon_{Bj}) 
\end{equation}  
\begin{equation}
\label{SU(3)}
 N_{u} + N_{d}  - 4 N_{{s}} =(3F-D)(1+\epsilon_{SU(3)}),
\end{equation}
and we take them as a measure of the
degree of fulfillment of the Bjorken sum rule \cite{BJ} and the $SU(3)$ symmetry.

Equations (\ref{Bj}) and (\ref{SU(3)}) allow to write the normalization of the three quark flavors in terms
of $N_{{s}}$, $\epsilon_{Bj}$, and $\epsilon_{SU(3)}$. Notice that no constraints have been imposed on the 
breaking parameters since we expect them to be fixed by data. The remaining parameters are constrained in such 
a way that positivity with respect to GRV98 parton distributions is fulfilled.
These last parameterizations are used in order to compute the denominators of equations (1) and (3).
This is particularly relevant 
at large $x$, and since no polarized data is available in that kinematical region, we directly fix the parameters
$\beta_u=3.2$, $\beta_d=4.05$ and $\beta_g=6$  for the NLO sets in agreement with GRV98.
Consistently with the choice for the unpolarized parton distributions, we use the values of $\Lambda_{QCD}$ 
given in Ref.\cite{GRV98} to compute $\alpha_s$ at NLO.

As antiquark densities we take 
\begin{equation}
 x\Delta \overline{q} =N_{\overline{q}}
\frac{x^{\alpha_{\overline{q}}}(1-x)^{\beta_{\overline{q}}}}
{B(\alpha_{\overline{q}}+1,\beta_{\overline{q}}+1)},
\end{equation}
for $\overline{u}$ and $\overline{d}$ quarks, and we assume $\overline{s}=s$
since the possibility of discrimination in the $s$ sector is beyond the
precision of the data (as in the unpolarized case).

Fragmentation functions are taken from \cite{KRETZER} and we also use the 
flavor separation criterion proposed there, which have shown to be in 
agreement with most recent analysis \cite{KE}. 

\begin{table}
\caption{\label{tab:table1} Inclusive and Semi-inclusive Data used in the fit. }
\begin{ruledtabular}
\begin{tabular}{ccccc} 
Collaboration & Target& Final state & \# points & Refs. \\ \hline
EMC          & proton& inclusive   &    10     & \cite{EMC} \\ 
SMC          & proton, deuteron & inclusive &  12, 12  & \cite{SMCi} \\  
E-143        & proton, deuteron & inclusive  &  82, 82  & \cite{E143} \\ 
E-155        & proton, deuteron & inclusive   &    24, 24    & \cite{E155} \\ 
Hermes       & proton,helium& inclusive   &    9, 9   & \cite{HERMES} \\
E-142        & helium& inclusive   &    8     & \cite{E142} \\ 
E-154        & helium& inclusive   &    17     & \cite{E143} \\  \hline 
SMC          & proton,deuteron& $h^+$, $h^-$  &  24, 24  & \cite{SMC} \\ 
Hermes       & proton, helium &  $h^+$, $h^-$   &    18, 18     & \cite{HERMES} \\  
\end{tabular}
\end{ruledtabular}
\end{table}

The data sets analyzed  include only points with $Q^2>1$ GeV$^2$
listed in Table 1,  and totaling 137, 118, and 34 points, from proton, deuteron, and
helium targets respectively, in the inclusive stage, plus 42, 24, and 18, from
proton, deuteron, and helium targets respectively, in the second stage.

Regarding the fitting procedure, this is done minimizing a function $f(\lambda_1,\lambda_2,...,\lambda_n)$ defined as 
\begin{equation}
f(\lambda_1,\lambda_2,...,\lambda_n)=\sum_i \lambda_i\,\chi^2_i,
\end{equation}
where the sum runs over the data sets or experiments $i$ included in the fit. The parameters $\lambda_i$ are the Lagrange multipliers defined for each data set and which can be varied in order to produce different fits where the relative weight of a given set of data has been modified. $\chi_i^2$ is the contribution to the overall $\chi^2$ coming from the data set $i$.

As it is well known, there are various alternatives for calculating these last contributions \cite{thornea}. The most simple and commonly used in fits to polarized data is adding the reported statistical and systematic errors in quadrature. This ignores the correlations between data points from the same measurements, but in many cases the full correlation matrices are not available.

As inclusive and semi-inclusive data are strongly correlated, and the correlation matrices are available, we have taken into account them, analyzing only  the inclusive data for SMC and Hermes corresponding to
`averaged' bins, for which the correlation matrices are defined. In the semi-inclusive case we only consider in the fit
the most precise data concerning the production of charged $\pm$ hadrons
(without identifying pions, kaons, or other particles individually). 

Regarding the normalization uncertainties, in reference \cite{DS} it was found that Hermes data was systematically below the best global fits, but allowing a floating normalization factor to it in order to account for the relative normalization uncertainties, the $\chi^2$ values were considerably improved. In the following we set allow this factor to be fixed by the global fit, finding the best fits for $12 \%$ correction.

\section{ Inclusive data}  

In this section we present results from the first stage of our analysis, which only deals 
with inclusive data. It is customary in NLO fits to inclusive data to present several 
sets of parton distributions with different alternatives for the features that are poorly 
constrained by the data, such as the gluon or strange sea quark polarization. As we are mainly concerned in assessing the goodness of the fit between theory and data
 and the degree of internal consistency of the data, rather than covering the different scenarios 
for parton densities, we just explore the more favored scenario of reference \cite{DS}, which was 
labeled as `set i' and has moderate gluon polarization. 

\begin{table}
\caption{\label{tab:table2} $\chi^2_i(\lambda)$ and $\chi^2_{Inclusive}(\lambda)$ for different data sets.}
\begin{ruledtabular}
\begin{tabular}{ccccccc} 
             & \multicolumn{2}{c}{$\lambda=1$}& \multicolumn{2}{c}{$\lambda=20$}& \multicolumn{2}{c}{$\lambda=100$} \\
            &$\chi^2_i$&$\chi^2_{Inc}$&$\chi^2_i$&$\chi^2_{Inc}$&$\chi^2_i$&$\chi^2_{Inc}$\\\hline
EMC-p        &    4.49  &            &     4.17 &   227.85   &    3.46   &  262.71      \\ 
SMC-p        &    3.84  &            &     3.32 &   225.21   &    3.20   &  226.71      \\ 
SMC-d        &   14.44  &            &    12.81 &   245.92   &   12.10   &  264.77      \\ 
E-143-p      &   60.70  &            &    60.22 &   224.83   &   60.21   &  224.95       \\
E-143-d      &   83.38  &            &    80.84 &   235.32   &   80.17   &  248.08       \\
E-142-n      &    4.68  &    224.59  &     2.40 &   233.52   &    1.51   &  251.98       \\ 
E-155-p      &   17.15  &            &    16.24 &   227.03   &   16.23   &  233.55       \\
E-155-d      &   17.10  &            &    16.89 &   225.65   &   14.98   &  277.42       \\
E-154-n      &    6.91  &            &     4.19 &   229.16   &    3.97   &  232.02       \\
Hermes-p     &    5.15  &            &     4.30 &   225.60   &    2.83   &  234.42       \\
Hermes-He    &    6.76  &            &     6.39 &   228.31   &    5.67   &  248.09       \\ 
\end{tabular}
\end{ruledtabular}
\end{table}

\setlength{\unitlength}{1.mm}
\begin{figure}[!]
\begin{picture}(100,90)(0,0)
\put(10,-15){\mbox{\epsfxsize8.9cm\epsffile{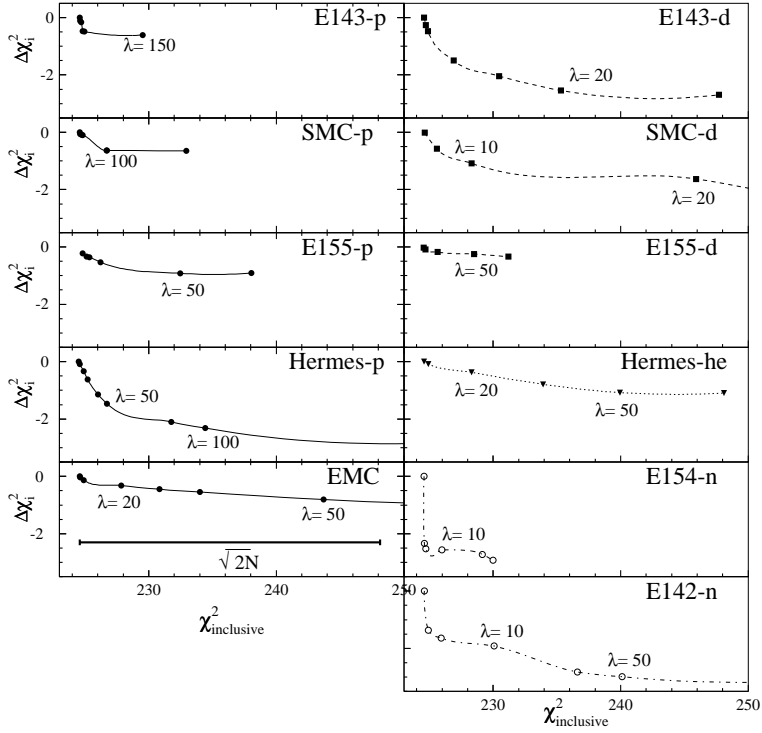}}}
\end{picture}
\caption{\label{fig:incchi}{\em 
 $\chi^2_{i}(\lambda)$ Against 
$\chi^2_{Inclusive}(\lambda)$ for all inclusive data sets.}}
\end{figure}

In Figure~\ref{fig:incchi} we show the outcome of different global fits to all the available 
inclusive data. The plot give the variation of the
$\chi^2_i$ value of each experiment against the total $\chi^2$
 value of the fit ($\chi^2_{Inclusive}$). The first point to the left of each curve 
($\lambda_i =1$) correspond to standard fits where no extra weight was given to none of the data sets. The parameters for this fit are presented in Table A1 in the appendix. Along the curves, the subsequent points come from fits where, following the Lagrange multiplier method explained in the previous section, increasing values of $\lambda_i$ have been given to a specific set of data, while keeping the others parameters $\lambda_j$ equal to 1.

The normal expectation in a good fit to data sets that individually only determine a small fraction 
of the parameters, is a monotonic decrease of a few units in the  $\chi^2_i$ of the subset 
which has been subject to the increase in weight while $\chi^2_{Inclusive}$ varies in the range 
$N\pm \sqrt{2N}$. One would also expect $\chi^2_i$ to
approach a saturation point within an increase of $\chi^2_{Inclusive}$ smaller than $\sqrt{2N}$.

As it can be seen in Figure~\ref{fig:incchi}, in all the cases, the curves 
show the features expected for consistent subsets of data, each one able to fix a limited number of 
parameters. The $\chi^2_{i}(\lambda)$ values of each experiment $i$ are shown in Table 2. 

\setlength{\unitlength}{1.mm}
\begin{figure}[!]
\begin{picture}(100,90)(0,0)
\put(2,-27){\mbox{\epsfxsize10.5cm\epsffile{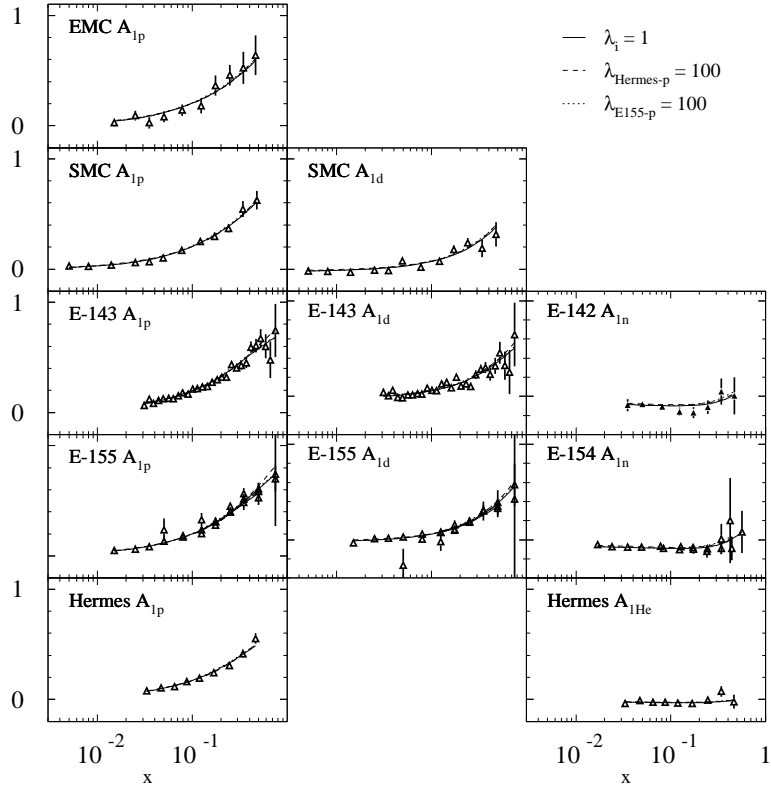}}}
\end{picture}
\caption{\label{fig:incfit}{\em  Fit to inclusive data ($\lambda=1$) 
together with the Hermes-p and E155-p driven fits with $\lambda=50$.}}
\end{figure}

Notice that the initial fall in the $\chi^2_{i}(\lambda)$ values with a very mild variation in 
$\chi^2_{Inclusive}$ illustrates the situation where a subset of data would be able to effectively 
fix some parameters in the fit, in general agreement with the remaining data, if it statistical 
significance were increased. This is clearly the situation of neutron and deuteron target data.

The standard $\lambda_i=1$ fit (solid line) can be seen in Figure~\ref{fig:incfit} together with the Hermes-p driven fit 
for $\lambda_{Hermes}=100$ (dashes) and that for E155 with  $\lambda_{E155}=100$ (dots) for comparison. As it can be noticed,
the changes in the asymmetries due to the extra weight in these subsets of data are almost negligible, as it 
can be expected from the moderate variations they produce in  $\chi^2_{Inclusive}(\lambda)$. For these reasons
we conclude that inclusive data is internally consistent and in excellent agreement with theory.

\section{Semi-Inclusive data}

In this section we focus on the consequences of including SIDIS data in the global fits. These data allow
in principle to discriminate between light sea quark flavors, so the corresponding parton distributions are
now parameterized and fitted. As we have anticipated, the inclusion of SIDIS data leads to acceptable fits 
according to the hypothesis testing criterion \cite{DS}, however the small statistical impact of SIDIS data
relative to DIS data hinders definite conclusions on sea quark distributions. 

A closer examination allowing for example increased weights in the different SIDIS data subsets, as we did 
for inclusive sets reveals some interesting features, as shown in Figure 3. Again, we have applied a 12 \% normalization factor to Hermes data, and in order to simplify the analysis, we consider SMC and Hermes proton target data together. Identical results are obtained if the data is discriminated for each experiment.

\setlength{\unitlength}{1.mm}
\begin{figure}[!]
\begin{picture}(100,85)(0,0)
\put(7,-20){\mbox{\epsfxsize9.5cm\epsffile{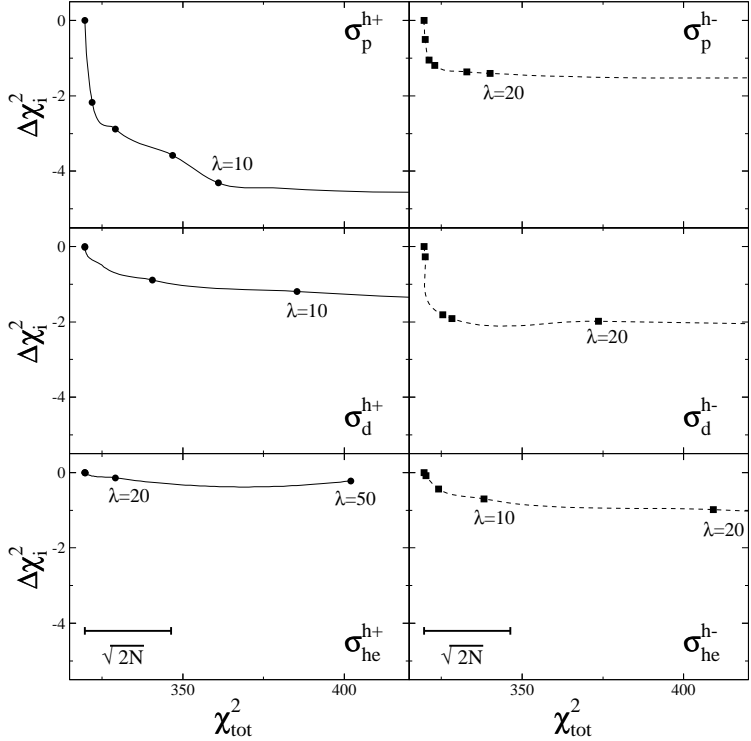}}}
\end{picture}
\caption{\label{fig:sincchi}{\em  $\chi^2_{i}(\lambda)$ against 
$\chi^2_{tot}(\lambda)$ for semi-inclusive data sets}}
\end{figure}

Although for most SIDIS data sets, $\chi^2_i(\lambda)$ reach their respective saturation values within a $\sqrt{2N}$
shift in $\chi^2_{tot}$ as required for overall consistency, data coming from positively charged 
hadroproduction on proton targets ($\sigma_p^{h+}$) seems to lay in the borderline, with a fall of several units and a saturation value outside de $\sqrt{2N}$ range. SIDIS data is mainly dominated
by proton target data but while $\sigma_p^{h-}$ driven fits lead to asymmetries in agreement with other
data sets, those driven by $\sigma_p^{h+}$ show an increasing disagreement with them, as can be seen in 
Figure~\ref{fig:sincfit}.

The inconsistency between these two data sets can also be seen in the parton distributions they produce, as shown in Figure ~\ref{fig:pdfmen} (a) and (b) , for $\sigma_p^{h-}$ and $\sigma_p^{h+}$ driven fits at $Q^2=5.0\, $ GeV$^2$, respectively. Notice that ($\Delta u + \Delta \overline{u}$) and ($\Delta d + \Delta \overline{d}$)
distributions, that should be fixed mainly by inclusive data, have minor, although not negligible,
changes in the fits driven by either data sets, showing the degree of consistency between inclusive and semi-inclusive data. However,
sea quark distributions depend strongly on which data set has received additional weight. The discrepancy 
is particularly strong for $\Delta \overline{u}$, which even change sign.

\setlength{\unitlength}{1.mm}
\begin{figure}[t]
\begin{picture}(0,95)(0,0)
\put(-50,-30){\mbox{\epsfxsize11.5cm\epsffile{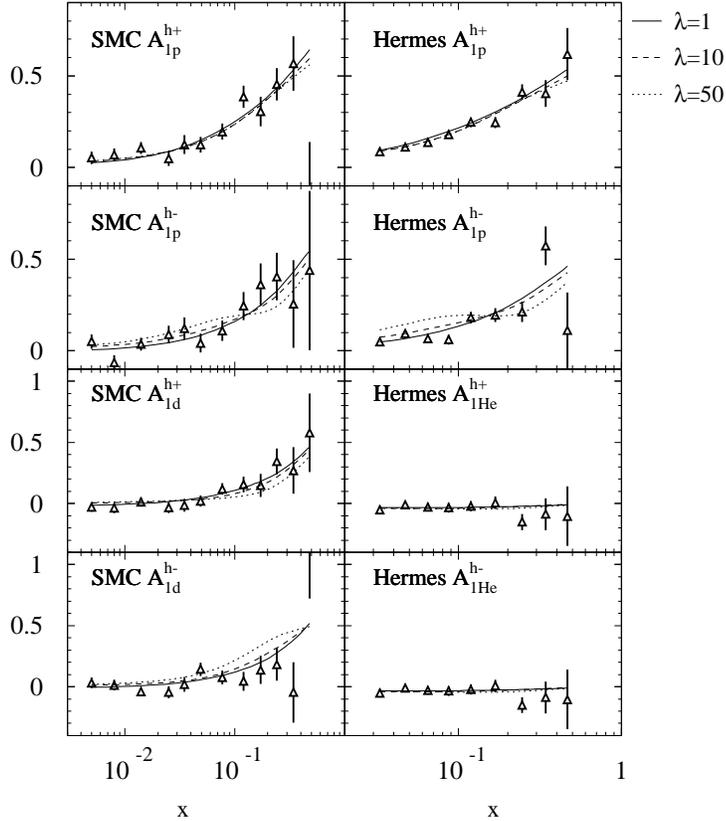}}}
\end{picture}
\caption{\label{fig:sincfit}{\em Fit to SIDIS data ($\lambda_i=1$) 
together with the $\sigma_p^{h+}$ driven fits. }}
\end{figure}

An interesting feature of SIDIS data is that for all subsets, the SIDIS-driven fits exceeds the 
$\sqrt{2N}$ range for $\chi^2_{tot}$ with values of $\lambda$ considerably smaller than the 
ones typical of inclusive data, as Figure~\ref{fig:sincchi} shows.  While inclusive data sets allows $\lambda_i$ values of 100 or more with a 
few units change in $\chi^2_{tot}$, SIDIS data exceeds the allowed range for $\lambda > 20$ or even less
in the case of $\sigma^{h+}$. In other words, at variance with what happens to DIS data, fits forced to reproduce SIDIS data lead to considerably poor global fits. This can be interpreted as a weaker level of consistency in the analysis 
of SIDIS data than in the inclusive case. In Table 3 we present $\chi^2_i$
values obtained for each data set in the standard fit ($\lambda_i=1$) and increasing the weights.

\setlength{\unitlength}{1.mm}
\begin{figure}[!]
\begin{picture}(-250,77)(0,0)
\put(-200,-10){\mbox{\epsfxsize7.0cm\epsffile{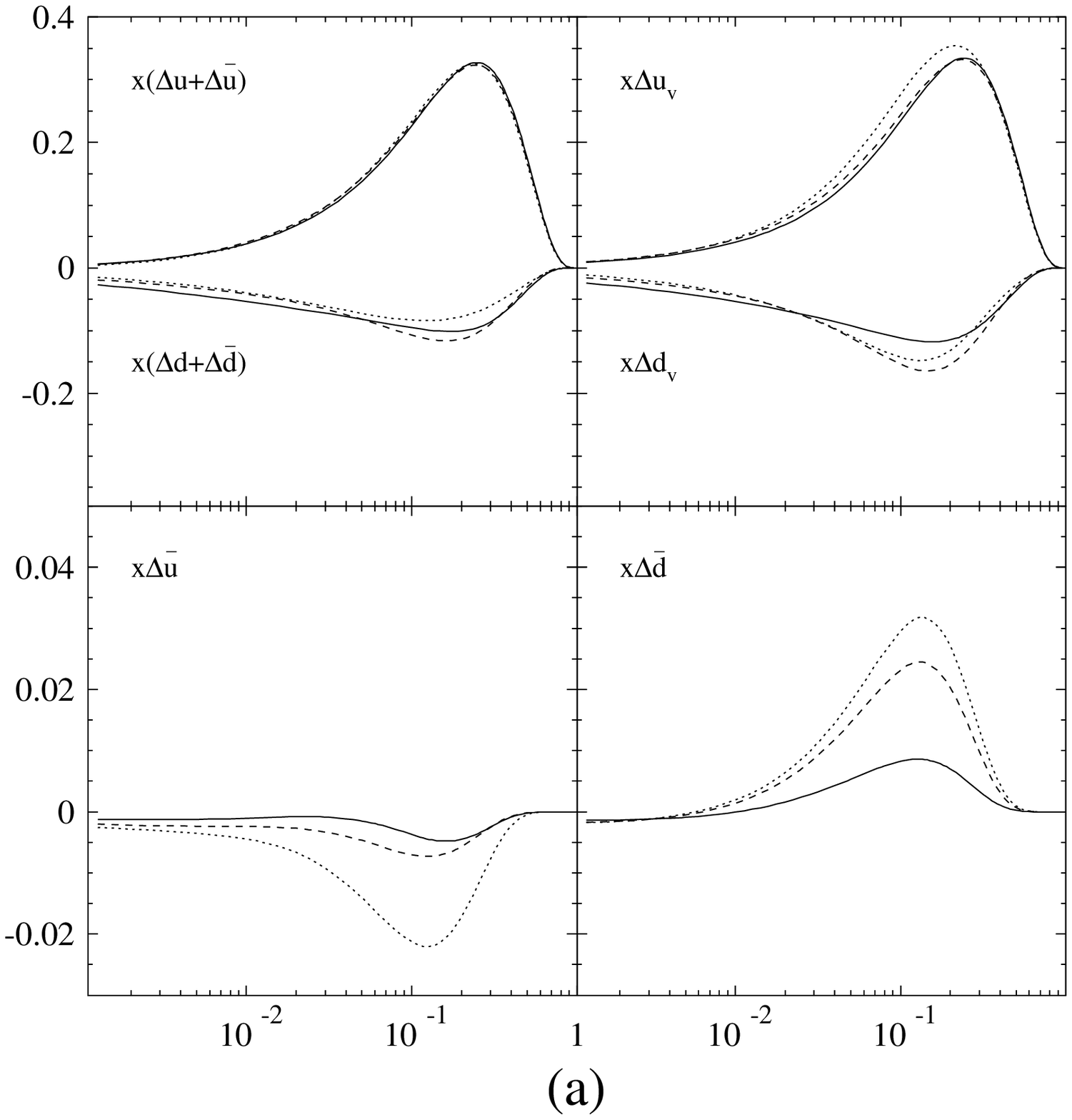}}}
\put(-120,-10){\mbox{\epsfxsize7.0cm\epsffile{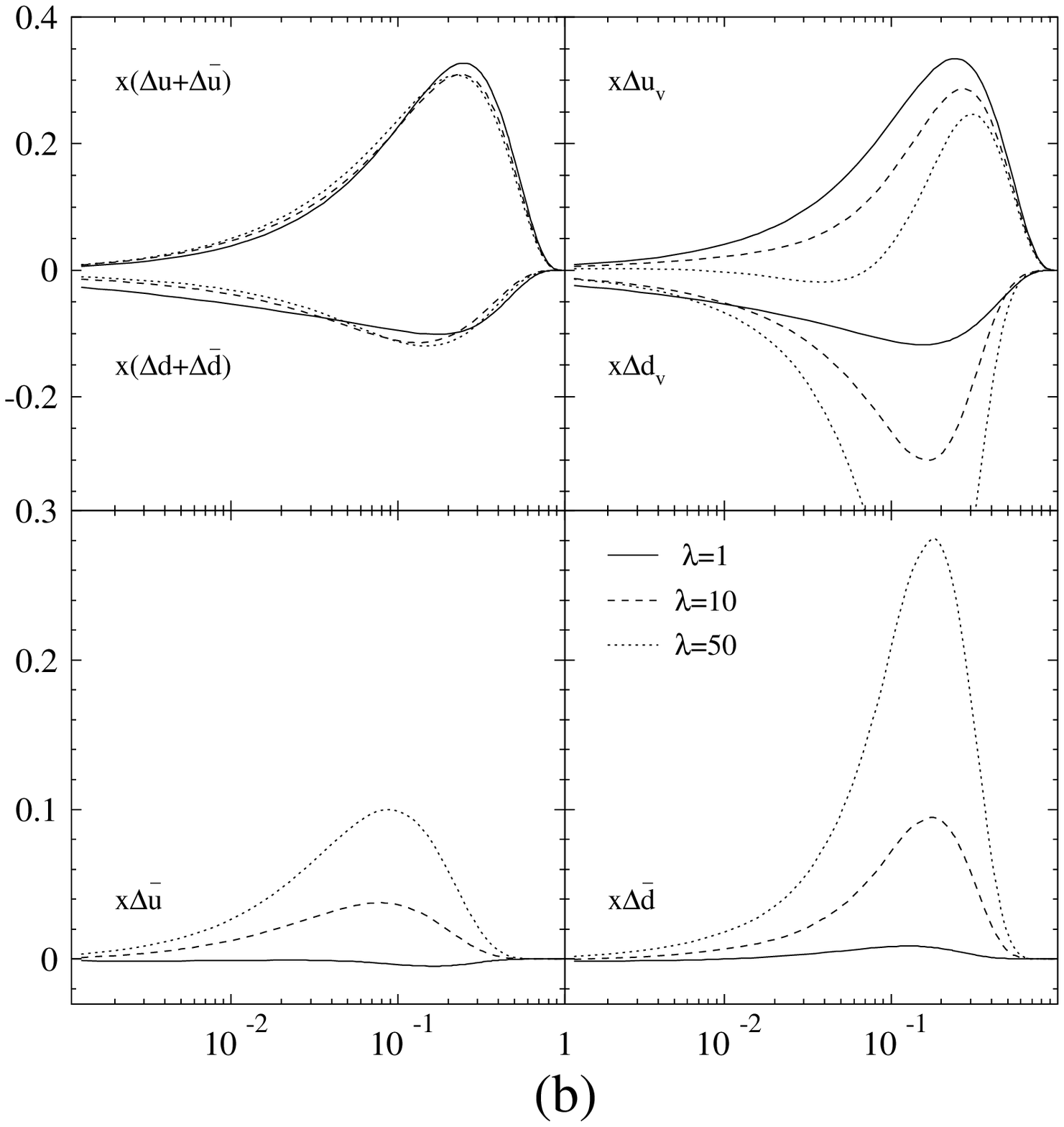}}}
\end{picture}
\caption{\label{fig:pdfmen}{\em Parton densities from $ (a)\,\sigma_p^{h-}$ and $ (b)\,\sigma_p^{h+}$ driven fits at $Q^2=5.0$ GeV$^2$. }}
\end{figure}  


In addition to the issue of the accuracy of the data, the analysis of SIDIS experiments relay also in our knowledge of unpolarized fragmentation functions. Although most of the uncertainties coming from these functions cancel out when computing asymmetries, the weaker degree of consistency, and odd behavior of $\sigma^{h+}$ regardless one uses proton or deuterium targets, and in different experiments, may hint
a failure in the extraction of fragmentation functions, particularly in the more troublesome discrimination between those for positive and negative final state hadrons.

\begin{table}
\caption{\label{tab:table3} $\chi^2_i(\lambda)$ and $\chi^2_{tot}(\lambda)$ for different SIDIS data 
sets.}
\begin{ruledtabular}
\begin{tabular}{ccccccc} 
             & \multicolumn{2}{c}{$\lambda=1$}& \multicolumn{2}{c}{$\lambda=10$}& \multicolumn{2}{c}{$\lambda=20$} \\
             &$\chi^2_i$&$\chi^2_{tot}$&$\chi^2_i$&$\chi^2_{tot}$&$\chi^2_i$&$\chi^2_{tot}$  \\ \hline 
$\sigma_p^{h+}$&   25.61     &            &   21.30  &   360.96   &    20.94  &450.20        \\ 
$\sigma_p^{h-}$&   26.93     &            &   25.57  &   332.89   &    25.53  &340.10        \\ 
$\sigma_d^{h+}$&    5.18     &  319.69    &    3.99  &   385.29   &     3.72  &504.24        \\ 
$\sigma_d^{h-}$&   14.71     &            &   12.80  &   328.27   &    12.73  &373.58        \\
$\sigma_{he}^{h+}$&    5.39  &            &    5.36  &   319.79   &     5.25  &329.17      \\ 
$\sigma_{he}^{h-}$&    6.69  &            &    6.46  &   338.18   &     5.71  &409.15     \\ 
\end{tabular}
\end{ruledtabular}
\end{table}

\section{Conclusions}

We have assessed the internal consistency in a NLO analysis of different sets of polarized DIS and SIDIS
data. For the inclusive data, the agreement shown between theory and data from the stand point of the 
hypothesis testing criterion, is confirmed when a detailed analysis using a variant of the parameter fitting criterion is performed.
For SIDIS data the level of consistency is considerably weaker, particularly in the case
of positively charged final state hadron data. Even though this kind of analysis can not establish 
whereas either data or some particular ingredient in the theoretical approach is responsible for the discrepancy, the unexpected features found in the analysis of $\sigma^{h+}$ data, for different targets and coming from different experiments, hint an inaccuracy in the separation between positively and negatively charged hadron fragmentation functions.   


 \subsection*{Acknowledgments}
We warmly acknowledge D. de Florian, A. Daleo and C. A. Garc\'{\i}a Canal for 
helpful comments and discussions.


\newpage
  
\setcounter{table}{0}
\def\tablename{Table A}
\section{Appendix: Parameters of the fit}
We present here the parameters of the fit with $\lambda =1$, for the first stage where we only deal with the inclusive data ($DIS\, Fit$) and the second stage, where the semi-inclusive data were included ($SIDIS\,Fit$).

\begin{table}[b]
\caption{\label{tab:tablea1} Parameters for the $lambda  =1$ fit.}
\begin{ruledtabular}
\begin{tabular}{ccc} 
\renewcommand{\arraystretch}{-1.5}
$Parameter$ & $DIS\, Fit$ & $SIDIS\, Fit$ \\ \hline
$\epsilon_{Bj}$       &    -0.002  &       -0.004  \\ \hline
$\epsilon_{SU(3)}$    &     0.085  &        0.088   \\ \hline
$\alpha_u$            &     0.858  &        0.858    \\ \hline
$\beta_u$             &     3.200  &        3.200        \\ \hline
$\gamma_u$            &    14.929  &       14.969         \\ \hline
$\delta_u$            &     1.004  &        1.005          \\ \hline
$\alpha_d$            &     0.434  &        0.433    \\ \hline
$\beta_d$             &     4.050  &        4.050        \\ \hline
$\gamma_d$            &    13.888  &       13.939         \\ \hline
$\delta_d$            &     1.651  &        1.651          \\ \hline
$N_s$                 &    -0.074  &       -0.075         \\ \hline
$\alpha_s$            &     2.500  &        2.491    \\ \hline
$\beta_s$             &    10.000  &       10.000         \\ \hline
$N_g$                 &     0.239  &        0.238         \\ \hline
$\alpha_g$            &     1.499  &        1.499    \\ \hline
$\beta_g$             &     6.000  &        6.000         \\ \hline
$N_{\bar u}$          &       -    &       -0.014         \\ \hline
$\alpha_{\bar u}$     &       -    &        2.311    \\ \hline
$\beta_{\bar u}$      &       -    &        7.646         \\ \hline
$N_{\bar d}$          &       -    &        0.014         \\ \hline
$\alpha_{\bar d}$     &       -    &        2.315    \\ \hline
$\beta_{\bar d}$      &       -    &        7.646         \\ 
\end{tabular}
\end{ruledtabular}
\end{table}

\end{document}